\documentclass[12pt]{article} 
\usepackage{amsmath}
\usepackage{amssymb}
\usepackage{epsfig}
\usepackage{fullpage}

\begin{document}
\centerline{\bf\Large Scattering phase shift and resonance properties}

\vspace{0.2cm}

\centerline{\bf\Large on the lattice: an introduction}

\vspace{1cm}

\centerline{\large S. Prelovsek$^{(a,b)}$*, C. B. Lang$^{(c)}$ and D. Mohler$^{(d)}$ \footnote{ Presented by S. Prelovsek at Bled Mini-Workshop 2011, {\it Understanding Hadronic spectra}, 3-10 July, Bled, Slovenia.  }}

\vspace{0.7cm}
{\small 
\centerline{(a) Jozef Stefan Institute, Jamova 39, 1000 Ljubljana, Slovenia}

\vspace{0.1cm}

\centerline{(b) Department of Physics, University of Ljubljana, Jadranska 19, 1000 Ljubljana, Slovenia}

\vspace{0.1cm}

\centerline{(c) Institut f\"{u}r Physik, FB Theoretische Physik, Universit\"{a}t Graz, A-8010 Graz\, Austria}

\vspace{0.1cm}

\centerline{(d) TRIUMF, 4004 Wesbrook Mall Vancouver, BC V6T 2A3, Canada}
}
\vspace{0.2cm}

\centerline{e-mail: sasa.prelovsek@ijs.si}

\vspace{1cm}

{\bf Abstract}

\vspace{0.2cm}

We describe the method for extracting the elastic scattering phase shift from a lattice simulation at an introductory level, for non-lattice practitioners. 
We consider the  scattering in a resonant channel, where the 
resulting phase shift $\delta(s)$ allows the  lattice determination of the mass and the width of the resonance from a Breit-Wigner type fit.  We present the method for the example of P-wave $\pi\pi$ scattering in the $\rho$ meson channel. 
 
\section{Introduction}

\begin{figure}[bt]
\begin{center}
\includegraphics*[width=0.45\textwidth,clip]{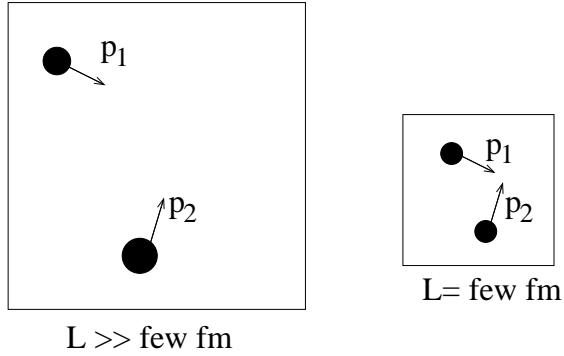}
\end{center}
\caption{ The energy of two hadrons in a box of size $L$. On the left, $L\gg $ fm and $E(L)\simeq E_{H1}(\mathbf{p_1})+E_{H2}(\mathbf{p_2})$. On the right, $L\simeq $ few fm and energy gets shifted due to their interaction, i.e. $E(L)\simeq E_{H1}(\mathbf{p_1})+E_{H2}(\mathbf{p_2})+\Delta E(L)$.   }\label{fig:two_mesons}
\end{figure}

The determination of the strong decay width of a hadronic resonance in lattice QCD is a much more demanding task than the determination of its approximate mass. The only available method (that was applied up to now) was proposed by L\"uscher \cite{luscher_phases} and is rather indirect. It applies for the case when the resonance appears in the elastic scattering of two hadrons $H_1H_2\to R \to H_1H_2$.  
\begin{itemize}
\item
First, the energy spectrum $E_n$ of the system  of two interacting hadrons 
$H_1H_2$ enclosed in a few-fermi box has to be determined. The system is illustrated in Fig. \ref{fig:two_mesons}.  
The spectrum in a finite box $E_n$ is discrete and few (one or two) lowest 
energy levels have to be determined by lattice simulation. 
\item
The shift of the energy $E_n$ with respect to the non-interacting energy
$E_{H1}(\mathbf{p_1})+E_{H2}(\mathbf{p_2})$
($E_{Hi}(\mathbf{p_i})=\sqrt{m_i^2+\mathbf{p_i}^2}$) gives info on the
interaction between $H_1$ and $H_2$. L\"uscher derived a rigorous relation
between the energy shift $E_n-E_{H1}-E_{H2}$ and the elastic phase shift
$\delta(s)$ for $H_1H_2$ scattering in continuum \cite{luscher_phases}. The measured energies
$E_n$ can be used to extract the phase shift $\delta(s)$ evaluated at
$s=E_n^2-\mathbf{P}^2$, where $E_n$ is the energy of the system and $\mathbf{P}$
its total momentum. In order to extract $\delta(s)$ at  several different values
of $s$, the simulations are done for several choices of total momenta
$\mathbf{P}$ of the $H_1H_2$ system, which leads to different values of  
$s=E_n^2-\mathbf{P}^2$. 
\item
The resulting dependence of $\delta(s)$ as a function  of $s$ can be used to extract the mass $m_R$ and the width $\Gamma_R$ of the resonance  $R$, which appears in the elastic channel $H_1H_2\to R \to H_1H_2$. For this purpose, 
the $\delta(s)$ can be fitted with a Breit-Wigner form or some other phenomenologically inspired form, which depend on $m_R$ and $\Gamma_R$. 
\end{itemize}

The described method, needed for the determination of the resonance width $\Gamma_R$, is rather challenging. It requires very accurate determination of a few lowest energy levels  of the system $H_1H_2$, since the resulting phase shift depends ultimately on the energy shift. Among all the meson resonances, this method has been up to now rigorously applied only to $\rho$ resonance. Although L\"uscher proposed the method already in late 80's \cite{luscher_phases}, 
the first lattice attempt to employ it to hadronic resonances had to wait until 2007 \cite{rho_cppacs}. Since then, several studies of $\rho$ have been carried out \cite{rho_qcdsf,rho_bmw}, with the most up to date ones  \cite{rho_etmc,rho_our,rho_pacscs}. 

This talk briefly describes the method to extract $\delta(s)$, $m_R$ and $\Gamma_R$ on an example of $\pi\pi$ scattering in the $\rho$ channel. It is  based on a recent simulation  \cite{rho_our}, 
which is the statistically most accurate determination of any strong meson width on one lattice ensemble. The purpose of this talk  is to  highlight the main physical reasoning, which lies behind the lattice extraction of $\delta(s)$, $m_R$ and $\Gamma_R$, omitting most of technical details. 
\vspace{0.2cm}

The sections follow the order of steps required, which are listed as items in
the introduction. Section II describes the determination of spectrum $E_n$ of
the coupled system $H_1H_2\leftrightarrow R$. The Section III described why 
$E_n$ allow one to extract the elastic phase shift $\delta(s)$. The extraction
of the resonance parameters $m_R$ and $\Gamma_R$ from the phase shift
$\delta(s)$ is done in Section IV.    We end with conclusions. 

\section{Spectrum of two hadrons in a finite box}

The $\rho$ meson is a resonance in $\pi\pi$ scattering in $P$-wave, and has quantum numbers $I^G(J^{PC})=1^+(1^{--})$. The total momentum $\mathbf{P}$ of the coupled   $\pi\pi-\rho$ system can have values $\tfrac{2\pi}{N_L}\mathbf{d}~,\ d\in Z^3$ due to the periodic boundary condition in the spatial direction, and we use the following three choices 
\begin{equation}
\label{P}
\mathbf{P}=(0,0,0)\ ,\ \tfrac{2\pi}{N_L}(0,0,1)\ ,\ \tfrac{2\pi}{N_L}(1,1,0)\quad {\mathrm{and\  permutations}}~.
\end{equation}
This enables us to obtain several values of $s=E_n^2-\mathbf{P}^2$ for the 
system, thereby allowing the
determination of $\delta(s)$ for these values of $s$ without changing
the spatial volume.  

Our simulation is performed on an ensemble of 280  
\cite{hasenfratz} gauge configurations with dynamical 
$u/d$ quarks, where the valence and dynamical quarks employ improved Wilson-Clover action. The corresponding pion mass is $m_\pi a=0.1673\pm 0.0016$ or $m_\pi=266\pm 4$ MeV. The lattice spacing is  $a=0.1239\pm 0.0013$ fm and we employ a rather small volume $N_L^3\times N_T=16^3\times 32$, which allows us to use the costly 
full distillation method \cite{peardon} for evaluating the quark contractions. 

\vspace{0.2cm}

On the lattice, the discrete energies of the system $E_n$ can be  extracted after computing the dependence of the  correlation matrix $C_{ij}(t_f,t_i)$ on Euclidean time $t_f-t_i$ 
\begin{equation}
\label{cor}
C_{ij}(t_f,t_i)=\langle 0|{\cal O}_i(t_f)~{\cal O}_j^\dagger(t_i)|0\rangle=\sum_n\langle {\cal O}_i|n\rangle \langle n|{\cal O}_j^\dagger\rangle ~e^{-E_n(t_f-t_i)}~. 
\end{equation}
The analytical expression on the right is obtained by inserting the complete set $\sum_n|n\rangle\langle n|$  of physical states $n$ with given quantum numbers. The interpolators $O_i$ have the quantum numbers of the system in question. In our case the interpolators have quantum numbers $J^{PC}=1^{--}$ and $|I,I_3\rangle=|1,0\rangle$ and total three-momentum $\mathbf{P}$. They have to couple well to 
 the $\pi\pi$ state and the quark-antiquark resonance $\rho$. 

For each choice of $\mathbf{P}$ (\ref{P}), we use 16 interpolators, listed in detail in Eq. (21) of \cite{rho_our}. 
We employ fifteen interpolators of quark-antiquark type  
\begin{equation}
\label{O_qq}
{\cal O}^{\bar qq}_i(t)=\sum_{\mathbf{x}}~e^{i\mathbf{Px}}~\tfrac{1}{\sqrt{2}}[\bar u {\cal F}_i u~(t,\mathbf{x})~+~\bar d {\cal F}_i d~(t,\mathbf{x})]~,
\end{equation}
 where ${\cal F}_i$ denotes different color-spin-space structures with the same resulting quantum number $J^{PC}=1^{--}$ and $|I,I_3\rangle=|1,0\rangle$.   We use also one $\pi(\mathbf{p_1})\pi(\mathbf{p_2})$ interpolator, where each pion is projected to a definite momentum  
\begin{equation}
\label{O_pipi}
{\cal O}^{\pi\pi}(t)=\tfrac{1}{\sqrt{2}}
[\pi^+(\mathbf{p_1})\pi^-(\mathbf{p_2})-\pi^-(\mathbf{p_1})\pi^+(\mathbf{p_2})]
\ ,\quad \mathbf{p_1}+\mathbf{p_2}=\mathbf{P}\, \quad \pi^{\pm}(\mathbf{p_i})=\sum_{\mathbf{x}}e^{i \mathbf{ p_i x}}~ \bar q\gamma_5 \tau^{\pm}  q~(t,\mathbf{x})
\end{equation}
In practice, the $\pi\pi$ interpolator is the most  important among our 16 interpolators, since it couples  to the scattering state much better than the quark-antiquark interpolators. Let us note that all other lattice studies aimed at $\Gamma_\rho$ used at most one quark-antiquark and one $\pi\pi$ interpolator, which may not always allow for reliable extraction of the first excited energy level $E_2$.    

\begin{figure}[bt]
\begin{center}
\includegraphics*[width=0.45\textwidth,clip]{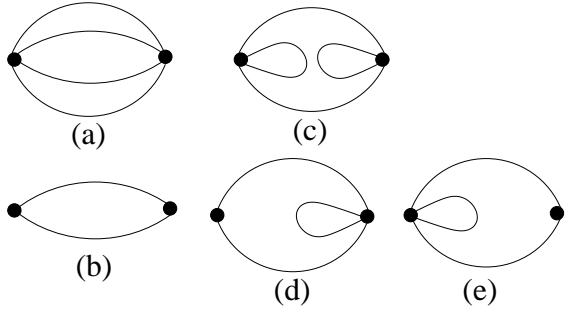}
\end{center}
\caption{ Contractions for $I=1$ correlators with  $\bar qq$ (\ref{O_qq}) and $\pi\pi$ (\ref{O_pipi}) 
interpolators.}\label{fig:contractions}
\end{figure}

Given the 16 interpolators, we compute the $16\times 16$ correlation matrix $C_{ij}(t_f,t_i)$ for all initial and final time-slices $t_i,t_f=1,..,N_T=32$. The needed Wick contractions that enter the correlation matrix with our $\bar qq$ and $\pi\pi$ interpolators are depicted in  Fig. \ref{fig:contractions}. The contributions (a,c,e) in Fig. \ref{fig:contractions} cannot be evaluated solely from the quark propagator from one point $(t_i,{\mathbf x_i})$ to all other points of the lattice (such a propagator   allowed most of the spectroscopy studies in the past). The contributions (a,c,e) require the propagators from all and to all points on the lattice, which is too costly to evaluate in practice. We use the recently proposed distillation method for this purpose \cite{peardon}, which enables the exact computation of the required contractions. 

We average the resulting correlators (i) over all initial time slices $t_i$ at fixed time separation $t_f-t_i$, (ii) over all directions of momenta ${\mathbf P}$ (\ref{P}) and (iii) over all directions of the $\rho$ meson polarization. 

The time dependence $t_f-t_i$ of the correlators $C_{ij}(t_f,t_i)$ (\ref{cor}) contains the information on the energies of the system $E_n$, and several methods for extracting $E_n$ from $C_{ij}$ are available. We extract two lowest energy levels $E_{n=1,2}$ of the system from the $16\times 16$ correlation matrix  $C_{ij}(t_f,t_i)$ using the  so called variational method \cite{gevp}, which is the most established among the available methods. Table \ref{tab:results}  displays the extracted lowest two energies $E_{n=1,2}$ of the coupled $\pi\pi-\rho$ system for our three choices of   total momenta ${\mathbf P}$ (\ref{P}).  

\begin{table*}[t]
\begin{center}
\begin{tabular}{c|cccc}
$\mathbf{P}$ & level $n$ &  $E_n\,a$ & $s\,a^2$ & $\delta$\\
\hline
$\tfrac{2\pi}{L}(0,0,0)$ &1& 0.5107(40)& 0.2608(41) & 130.56(1.37)\\
$\tfrac{2\pi}{L}(0,0,0)$ &2& 0.9002(101)& 0.8103(182) & 146.03 (6.58) [*]\\
\hline
$\tfrac{2\pi}{L}(0,0,1)$ &1& 0.5517(26)  & 0.1579(29) &  3.06 (0.06)\\
$\tfrac{2\pi}{L}(0,0,1)$ &2& 0.6845(49)  & 0.3260(69) &156.41(1.56)\\
\hline
$\tfrac{2\pi}{L}(1,1,0)$ &1& 0.6933(33)  & 0.1926(49) & 6.87(0.38)\\
$\tfrac{2\pi}{L}(1,1,0)$ &2& 0.7868(116)& 0.3375(191) & 164.25(3.53)\\
\hline
\end{tabular}
\caption{\label{tab:results} The results for two lowest  levels $n=1,2$ of the coupled $\pi\pi-\rho$ system with three choices of total momentum $\mathbf{P}$ on our  lattice with $m_\pi a=0.1673\pm 0.0016$, $L=16a$ and the lattice spacing $a=0.1239\pm 0.0013$ fm. The energy levels $E_n$ are obtained by multiplying $E_na$ with $a^{-1}\simeq 1.6$ GeV. The invariant mass squared  of the system is $s=E_n^2-\mathbf{P}^2$, but the dimensionless  value in the table $s\,a^2$ is obtained using the discretized version of this relation  \cite{rho_our}.  
 }
\end{center}
\end{table*}

The spectrum $E_n$ in Table \ref{tab:results} for our finite box is the main
result of this section.  Each energy level corresponds to a different value of
$s=E_n^2-{\mathbf P}^2$, as calculated from $E_n$ and ${\mathbf P}$ in the Table
\ref{tab:results}. In fact, the table lists values of $s$ obtained from the
discrete lattice
version of the dispersion relation, which takes into account part of the corrections
to  $s=E_n^2-{\mathbf P}^2$ due to finite lattice spacing  \cite{rho_our}.

\section{Extraction of the phase shifts from energy levels}  

Let us consider the case when the resonance $R$ can strongly decay only to two spinless hadrons  $H_1$ and $H_2$, so one has elastic scattering of $H_1$ and $H_2$. We point out that the non-elastic case, when a resonance can decay strongly to several final states (i.e. $H_1H_2$ and $H_1'H_2'$), is much more challenging for a lattice study.  

Suppose one encloses two hadrons $H_1(p_1)~H_2(p_2)$ with three-momenta $p_1$ and $p_2$ 
into a large box of size $L\gg $ fm and measures their energy. In a large box, they hardly interact and their 
energy is equal to sum of individual energies $E^{non-int}=E_{H1}(p_1)+E_{H2}(p_2)$ with $E_{H}(p)=\sqrt{m_H^2+p^2}$. Now, let's force $H_1(p_1)$ and $H_2(p_2)$ to interact by 
decreasing the size of the box to 
 $L$ of a few fm. The energy of the system $E(L)=E_{H1}(p_1)+E_{H2}(p_2)+\Delta E(L)$ is shifted with respect to $E^{non-int}$: it will increase 
 ($\Delta E(L)>0$) if the interaction is repulsive and decrease ($\Delta E(L)<0$) if the interaction is attractive. 
   This simple physical reasoning indicates that the energy shift $\Delta E(L)$ gives info on the interaction. 
   
   \begin{figure}[tbh!]
\begin{center}
  \centerline{\includegraphics[height=5cm]{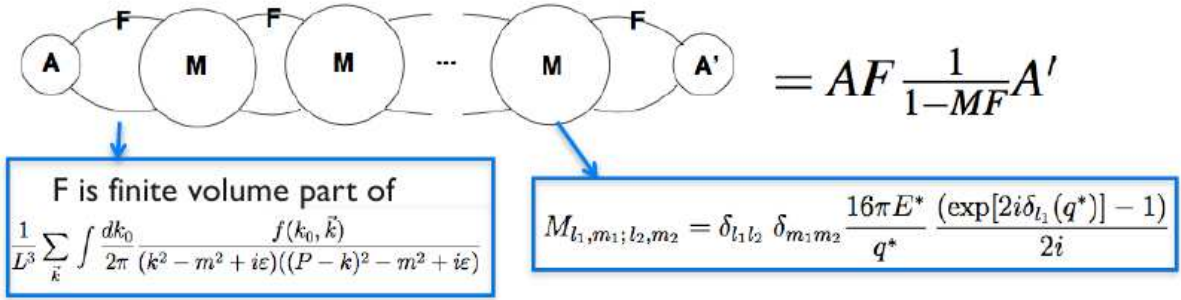}}
\caption{ The scattering of two interacting particles as series of the interaction vertex $M(\delta_L)$ and the scattering of non-interacting particles $F$ at finite $L$ \cite{sharpe}. }\label{sharpe}
\end{center}
\end{figure}
   
   In fact, the energy shift $\Delta E(L)$ and the energy itself $E(L)$ do not only give us "some" info on the interaction. According to the seminal analytic work of L\"uscher \cite{luscher_phases}, $E(L)$ or $\Delta E(L)$ 
   rigorously tells us the value of the elastic scattering phase shift of $H_1H_2$ scattering at $L\to \infty$, i.e. $\delta(L=\infty)$:
   \begin{equation}
  {\mathrm{Luscher\ method}}: \qquad E(L) \longrightarrow \  \delta(s,L=\infty)\qquad s=E(L)^2-\mathbf{P}^2
   \end{equation} 
    The derivation and the resulting formulae between $E(L)$ and $\delta$ are lengthy and rather complicated, 
    but let us briefly explain  at least why $E(L)$ contains info on $\delta(L=\infty)$. A nice and clear quantum-filed theory derivation is given in \cite{sharpe} and the main message is illustrated in Fig. \ref{sharpe}. 
    The scattering of two interacting spin-less hadrons $H_1H_2$ at finite $L$ (for degenerate case $m_{H1}=m_{H2}=m$) 
    is represented  in QFT by series of:
    \begin{itemize}
    \item 
    scattering of two non-interacting hadrons at finite $L$, represented by F.
    The expression $F$ contains  sums over the loop momenta $\mathbf{k}$, which are
    allowed in a finite box $L$ with periodic boundary conditions in space. Here
    $f(k_0,\mathbf{k})$ stands for dependence of the vertices on the left and
    right on $k_0$ and $\mathbf{k}$. 
    \item 
    the interaction vertex $M$ with four hadron legs. This vertex depends on the elastic phase shift 
    $\delta_l$ (at infinite volume) for the case of elastic scattering in the $l$-th partial wave. 
    \end{itemize}
     The physical scattering requires resummation of the bubbles in Fig. \ref{sharpe}, with non-interacting parts $F$ and the interacting parts $M$, giving  $AF\tfrac{1}{1-MF}A'$. The positions of the poles of the sum $AF\tfrac{1}{1-MF}A'$ obviously depend on $M$ and therefore on $\delta_l$. The positions of the poles dictate the possible energy levels of the system $E_n(L)$, so the energy levels $E_n(L)$ depend on $M$ and therefore on $\delta_l$.  
     
     The purpose of the above illustration was just to indicate why $E_n(L)$ depend on $\delta_l$. 
     In the case of $\pi\pi$ with $J^P=1^-$, the relevant wave has $l=1$ and we denote the corresponding phase by $\delta\equiv\delta_1$.  
     The complete analytic relations between $E_n(L)$ and $\delta(s)$ needed for our case of the 
     $\pi\pi$ scattering with $J^{PC}=1^{--}$ and $I=1$ are provided in \cite{rho_our} 
     (for every $|\mathbf{P}|$ a different form of relation applies).  These allow to extract $\delta$ for each of our  
     six energy levels in Table \ref{tab:results} and the resulting phase shifts are given in the same Table. 
     
     The presented L\"uscher formalism applies only for the case of elastic
     scattering.  The $\pi\pi$ state is the only scattering state in this
     channel for energies when $4\pi$ state cannot be created, i.e., when
     $s=E_n^2<(4m_\pi)^2$. For our $m_\pi a=0.1673$ this is valid for all six
     levels, with the exception of the level  $E_1$ at ${\mathbf P}=0$, which is
     above $4\pi$ inelastic threshold. As the L\"uscher analysis is not valid
     above the inelastic threshold, we omit this level from further analysis.
     
\begin{figure}[bt]
\begin{center}
  \centerline{\includegraphics[height=5cm]{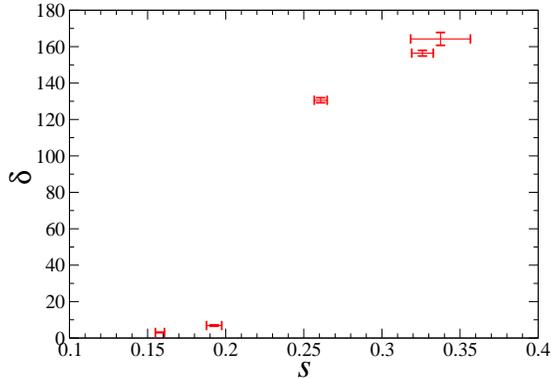}}
\caption{ The $\pi\pi$ phase shift $\delta(s)$ (in degrees) for five different values of dimensionless $sa^2=(E_na)^2-(\mathbf{P}a)^2$, extracted from our lattice study \cite{rho_our}. The $s$ is obtained by multiplying $sa^2$ with $(a^{-1})^2\simeq (1.6~GeV)^2$.   }\label{fig:phase_noline}
\end{center}
\end{figure}

     The resulting scattering phase shifts for five values of $s$ are shown in Fig. \ref{fig:phase_noline}. This is the main result of the lattice study; the resonance properties will be obtained by fitting $\delta(s)$ in the next section. 
     
     Note that the resulting phases are determined with a relatively good precision, which is better than in other 
     available lattice studies of $\rho$ at comparable $u/d$ quark masses. 
     The good precision can be traced back to 
     various advanced techniques we used: the distillation method for evaluating contractions, usage of a large interpolator basis and average over all initial time slices, directions of momenta $\mathbf{P}$ and polarizations of $\rho$.

\section{Extracting resonance mass and width from the  phase shift}

The phase shift $\delta(s)$ in Fig. \ref{fig:phase_noline}, obtained directly from the lattice study, can be used to extract the properties of the resonance, in our case the $\rho$. The phase shift has a typical resonance shape: it passes from $\delta\simeq 0^\circ$ to $\delta \simeq 180^\circ$: the point where it crosses $90^\circ$ gives the position of the resonance ($s=m_\rho^2$), while the steepness of the rise gives its width $\Gamma_\rho$. In particular, $\delta$ is related to resonance parameters by expressing the scattering amplitude $a_l$ in terms of $\delta$ on one hand, and with Breit-Wigner form in the vicinity of the resonance on the other hand
\begin{equation}
\label{amplitude}
a_1=\frac{-\sqrt{s}\,\Gamma(s)}{s-m_\rho^2+i \sqrt{s}\,\Gamma(s)}=
\frac{e^{2i \delta(s)}-1}{2i}~.
\end{equation}
Relation (\ref{amplitude}) can be conveniently re-written  as 
\begin{equation}
\label{amplitude1}
\sqrt{s}\,\Gamma(s)\,\cot \delta(s)=m_\rho^2-s~. 
\end{equation}
The decay width significantly depends on the phase space and therefore on $m_\pi$, so the decay width extracted 
at $m_\pi\simeq 266$ MeV could not be directly compared to the measured width. So, it is customary to 
extract the $\rho\to\pi\pi$ coupling $g_{\rho\pi\pi}$ instead of the width, where the width 
\begin{equation}
\label{width}
\Gamma(s)=\frac{{p^*}^3}{s} \frac{g_{\rho\pi\pi}^2}{6\pi}~,\qquad\Gamma_\rho=\Gamma(m_\rho^2)
\end{equation}
depends  on the phase space for a $P$-wave decay and the  coupling $g_{\rho\pi\pi}$. The coupling is expected to be only mildly dependent    on $m_\pi$, which was explicitly confirmed in the lattice studies \cite{rho_etmc,rho_pacscs}  and analytic study \cite{pelaez_mpi_dep}. In (\ref{width}), $p^*$ denotes the pion momentum in the center-of-momentum frame and we extract it from $s$ using a discretized version of relation $\sqrt{s}=2\sqrt{m_\pi^2+p^{*2}}$ \cite{rho_our}. Inserting  $\Gamma(s)$ (\ref{width}) into (\ref{amplitude1}), 
one obtains an expression for $\delta(s)$ in terms of two unknown parameters: $m_\rho$ and $g_{\rho\pi\pi}$. We fit these two parameters using five values of $\delta(s)$ given in Fig. \ref{fig:phase_noline} and Table \ref{tab:results}, and we get the values of resonance parameters in Table \ref{tab:results_resonance} with small statistical errors. 

\begin{figure}[bt]
\begin{center}
  \centerline{\includegraphics[height=5cm]{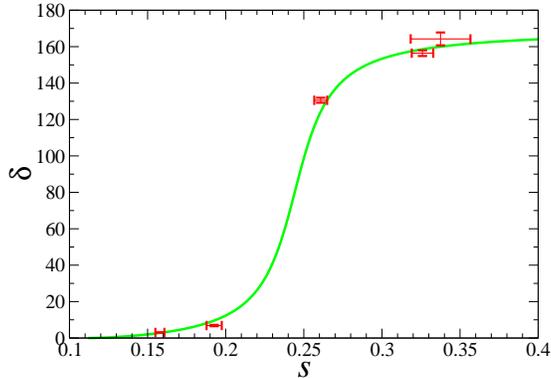}}
\caption{ The crosses are the $\pi\pi$ phase shift $\delta(s)$ (in degrees) for five
different values of dimensionless $sa^2=(E_na)^2-(\mathbf{P}a)^2$, extracted
from our lattice study \cite{rho_our}. The line is the Breit-Wigner fit
(\ref{amplitude1},\ref{width}) for the resulting $m_\rho$ and $g_{\rho\pi\pi}$
in Table \ref{tab:results_resonance}.  The physical value of $s$ is obtained by multiplying $sa^2$
with $(a^{-1})^2\simeq (1.6~GeV)^2$.  }\label{fig:phase_shift}
\end{center}
\end{figure}

The resulting $\rho$-meson mass in Table \ref{tab:results_resonance} is slightly higher than in experiment, as expected due to $m_\pi=266~MeV>m_\pi^{exp}$. The coupling  $g_{\rho\pi\pi}$ is rather close to the value $g_{\rho\pi\pi}^{exp}$ derived from the experimental width $\Gamma_\rho^{exp}$. 

\begin{table*}[t]
\begin{center}
\begin{tabular}{c | c c}
 &  lattice (this work \cite{rho_our}) & exp [PDG]\\
 & $m_\pi\simeq 266$ MeV & \\
\hline
 $m_\rho$& $792\pm 12$ MeV & $775$ MeV \\
$g_{\rho\pi\pi}$ & $5.13\pm 0.20$ & $5.97$ \\
\hline
\end{tabular}
\caption{Our lattice results for the resonance parameters \cite{rho_our}, compared to the experimental values. \label{tab:results_resonance}  
 }
\end{center}
\end{table*}

\section{Comparison to other lattice  and analytical studies}

The comparison of our results for $m_\rho$ and $\Gamma_\rho$  to  two recent lattice studies \cite{rho_etmc,rho_pacscs}  is compiled in Fig. 8 of  \cite{rho_pacscs}. Our result has the smallest error on a given ensemble, demonstrating  that accurate lattice determination $m_R$ and $\Gamma_R$ for (some) resonances is possible now. The other two lattice studies are done for two \cite{rho_pacscs} and four \cite{rho_etmc} 
pion masses and explicitly demonstrate mild dependence of $g_{\rho\pi\pi}$ on $m_\pi$. The discussion concerning the (dis)agreement of the three lattice studies is given in \cite{rho_pacscs} and will be extended in \cite{rho_our_pos}.

\begin{figure}[tbh!]
\begin{center}
  \centerline{\includegraphics[height=5cm]{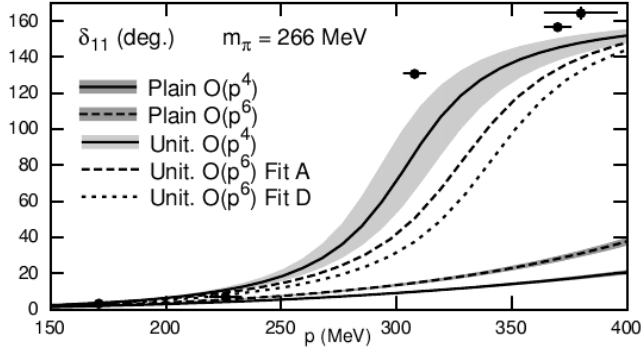}}
\caption{ The $\pi\pi$ phase shift in the $\rho$ channel $\delta_{11}(p)\equiv \delta(p^*)$ at $m_\pi=266$ MeV: the solid line (indicated by ``Unit $O(p^4)$'') gives prediction of the lowest order of Unitarized Chiral Perturbation Theory \cite{pelaez_comparison,pelaez_comparison_proc}, while bullets are our lattice data.   }\label{fig:comparison_pelaez}
\end{center}
\end{figure}

The comparison of our $\delta(s)$ to the prediction of the lowest non-trivial order of unitarized Chiral Perturbation Theory \cite{pelaez_comparison} is given by the solid line in \ref{fig:comparison_pelaez}, which has been recalculated for our $m_\pi=266$ MeV in \cite{pelaez_comparison_proc}. The lowest\footnote{One cannot make a fair comparison between out lattice result and the next-to-lowest order prediction, since it depends on a number of LECs, and some of them have been fixed using $m_\rho$ from another lattice study, which gets a significantly higher $m_\rho$.}
order prediction does not depend on unknown LECs and agrees reasonably well with our lattice result, 
given by the bullets.

\section{Conclusions}

We highlighted the main physical reasoning, which lies behind the lattice extraction of elastic phase shifts $\delta(s)$  and the resonance parameters $m_R$ and $\Gamma_R$. The purpose was to present the general principle of the method and omit the technical details. The method was presented on the example of $\pi\pi\to\rho\to \pi\pi$ scattering. This example demonstrates that a proper first-principle treatment of some hadronic resonances on the lattice is now possible.

\vspace{0.5cm}

{\bf Acknowledgments}

{\small 
We would like to kindly thank Anna Hasenfratz for providing the
gauge configurations used for this work.  We would like to thank Xu
Feng, Naruhito Ishizuka, Jose Pelaez, Gerrit Schierholz and Richard Woloshyn 
for valuable discussions. The calculations have been performed on the theory
cluster at TRIUMF and on local clusters at the University of Graz and
Ljubljana.  We thank these institutions for providing support. This work is
supported by the Slovenian Research Agency, by the European RTN network FLAVIAnet
(contract number MRTN-CT-035482) and by the Natural Sciences and Engineering
Research Council of Canada (NSERC).}

\end{document}